\documentclass[prd,twocolumn,nofootinbib,amsfonts]{revtex4}
\usepackage{graphicx,bm,epsfig,amsmath}
\newcommand{\be}{\begin{equation}}
\newcommand{\ee}{\end{equation}}
\newcommand{\Ham}{\mathcal{H}}
\newcommand{\Lag}{\mathcal{L}}
\newcommand{\Phimas}{\Phi^\dagger}
\newcommand{\Pimas}{\Pi^\dagger}
\newcommand{\vx}{\mathbf{x}}
\newcommand{\vp}{\mathbf{p}}
\newcommand{\deltat}{\delta^{(3)}}

\newcommand{\ena}{E_a}
\newcommand{\eb}{E_b}
\newcommand{\omp}{\omega_{\mathbf{p}}}
\newcommand{\intxt}{\int d^3\vx}
\newcommand{\intxc}{\int d^4 x}
\newcommand{\intpt}{\int \frac{d^3\vp}{(2\pi)^3}}
\newcommand{\intpc}{\int \frac{d^4 p}{(2\pi)^4}}
\newcommand{\udp}{U_{\vp}}
\newcommand{\vdp}{V_{\vp}}
\newcommand{\intauxx}{\int d^4 \tilde{x}}

\newcommand{\gaux}{\sqrt{\tilde{g}(\tilde{x})}}

\newcommand{\Lie}{\pounds_n}

\begin{document}

\title{The Theory of a Quantum Noncanonical Field in Curved Spacetimes}
\author{Javier Indur\'ain}
\email{indurain@unizar.es}
\affiliation{Departamento de F\'{\i}sica Te\'orica,
Universidad de Zaragoza, Zaragoza 50009, Spain}

\author{Stefano Liberati}
\email{liberati@sissa.it}
\affiliation{SISSA, Via Beirut 2-4, 34151 Trieste, Italy and INFN sezione di Trieste}

\begin{abstract}
Much attention has been recently devoted to the possibility that quantum gravity effects could lead to departures from Special Relativity in the form of a deformed Poincar\`e algebra. These proposals go generically under the name of Doubly or Deformed Special Relativity (DSR). In this article we further explore a recently proposed class of quantum field theories, involving noncanonically commuting complex scalar fields, which have been shown to entail a DSR-like symmetry.
An open issue for such theories is whether the DSR-like symmetry has to be taken as a physically relevant symmetry, or if in fact the ``true'' symmetries of the theory are just rotations and translations while boost invariance has to be considered broken.
We analyze here this issue by extending the known results to curved spacetime under both of the previous assumptions. We show that if the symmetry of the free theory is taken to be a DSR-like realization of the Poincar\'e symmetry, then it is not possible to render such a symmetry a gauge symmetry of the curved physical spacetime. However, it is possible to introduce an auxiliary spacetime which allows to describe the theory as a standard quantum field theory in curved spacetime. Alternatively, taking the point of view that the noncanonical commutation of the fields actually implies a breakdown of boost invariance,  the physical spacetime manifold has to be foliated in surfaces of simultaneity and the field theory can be coupled to gravity by making use of the Arnowitt-Deser-Misner prescription.
\end{abstract}

\maketitle

\section{Introduction}

Quantum gravity has historically suffered from  lack of observational support and problems with conceptual issues  common to all of the most studied proposals \cite{strings,lqg} (e.g.\/  the ``problem of time" and the ``background-independence problem" \cite{stachashtbook}). However, over the past few years there has been a growing interest in possible low energy, observable, effects of quantum gravity scenarios, ranging from TeV-scale quantum gravity to high energy departures from Local Lorentz invariance of spacetime (see e.g.\/ the discussion in Ref.~\cite{Mattingly,AmelinoCamelia:2008qg}). In particular, it was recently suggested that the Planck length $L_p$ ($L_p \sim 10^{-33} \text{cm} $) should be taken into account in describing the rotation/boost transformations between inertial observers.

In alternative to the standard approach of considering Planck suppressed Lorentz (and possibly CPT) invariance violations in effective field theory (EFT) (see e.g. Ref.~\cite{Mattingly,Mattingly:2007zz}), it was conjectured that the relativity principle could be preserved via a deformation of special relativity, in the form of the so-called ``Doubly or Deformed special relativity" (DSR) \cite{AmelinoCamelia:2000mn, AmelinoCamelia:2002wr, Judes:2002bw}. While this proposal is formulated in momentum space, a coordinate space definitive formulation is still lacking (and several open issues are still unresolved, see e.g.\/ Ref.~\cite{Un-Sch,Liberati:2004ju}).

Although some interpretation of DSR in commutative spacetime have been proposed \cite{Liberati:2004ju,AGGLLM},  there has been a growing activity \cite{KowalskiGlikman:2002we} toward connecting it  with noncommutative spacetime \cite{Majid:1994cy, Lukierski:1993wx, KowalskiGlikman:2004qa}, which also arise in 3d-Quantum Gravity \cite{Freidel:2005me} and String Theories \cite{connes}.
However, the connection between quantum gravity and noncommutative spacetime is in principle problematic: in fact the latter seem to entail a ``background-independence problem" given that the commutation relations among spacetime coordinates promote them to the role of {operators}, despite they should be treated as mere labels in a background independent theory (see e.g.\/ Ref.~\cite{Westman:2007yx}).

An alternative to modify the commutation relations between coordinates, while preserving background independence, is to modify the commutation relations of field operators on a spacetime manifold. If the manifold is flat spacetime, the Quantum Theory of Noncanonical Fields (QNCFT) has been studied \cite{Carmona:2002iv, Carmona:2003kh}\footnote{{These references named the theory Quantum Theory of Noncommutative Fields, but Quantum Theory of Noncanonical Fields provides indeed a more precise description.}} and it can be seen as a natural extension to field theory of Non Commutative Quantum Mechanics \cite{NCQM}.  Noticeably, it has been shown that there is a link between the one particle sector of QNCFT in flat spacetime and DSR~\cite{NCFisDSR} .

Here we are going to use the results of Ref.~\cite{NCFisDSR} to study the case of the QNCFT in Curved Spacetimes. In the Section II we will review the main results of Ref.~\cite{Carmona:2003kh}. In the Section III we will discuss the connection between QNCFT and DSR as derived in Ref.~\cite{NCFisDSR}. In the Section IV we will derive the Lagrangian, the equation of motion, and the internal product of solutions of the QNCFT in Curved Spacetimes in different cases depending on {which are the} symmetries of the free theory in flat spacetime that we want to promote to gauge symmetries of the curved spacetime. The Section V is devoted to some closing remarks.

\section{QNCFT in Flat Spacetime}

Let us consider the theory of a complex scalar field, given by the Hamiltonian density:
\be
\Ham = \Pimas(\vx)\Pi(\vx)+\bm{\nabla}\Phimas(\vx)\cdot\bm{\nabla}\Phi(\vx)+m^2\Phimas(\vx)\Phi(\vx) \label{ham1} \, ,
\ee
the Hamiltonian being the integral to spacelike coordinates of $\Ham$.
The field is quantized by imposing the noncanonical commutation relations:
\begin{eqnarray}
	\left[\Phi(\vx),\Phimas(\vx')\right] & = & \theta \deltat(\vx-\vx') \nonumber\, , \\
	\left[\Phi(\vx),\Pimas(\vx')\right] & = & i \deltat(\vx-\vx') \label{com1}\, ,\\
	\left[\Pi(\vx),\Pimas(\vx')\right] & = & B \deltat(\vx-\vx') \nonumber \, ,
\end{eqnarray}
where $\theta$ and $B$ are considered to be ``sufficiently small", so that the canonical quantum field theory should be seen as a good approximation to this theory for a certain range of energies or momenta.  It should be noticed that $1/\theta$ plays the role of an UV scale and $B$ plays the role of an IR scale \cite{Carmona:2003kh}.
We will forget for the moment about the IR scale and set $B=0$ in our calculations (the effect of the scale $B$ turns out to be, when $\theta = 0$, a redefinition of the mass $m$ and a constant energy gap between n-particle and (n+1)-particle states)~\cite{Carmona:2003kh}.
We shall see now that the just given set of definitions completely specifies the theory.

Let us start by noticing that the momentum operator
\be
\mathbf{P}=\intxt \left[\Pimas(\vx)\bm{\nabla}\left(\Phi(\vx)-\frac{i\theta}2\Pi(\vx)\right)+ \,h.c.\,\right]\, ,
\ee
commutes with the Hamiltonian, and hence the field can be expanded in terms of creation-annihilation operators of particles and antiparticles with a given momentum.
\begin{eqnarray}
\Phi(x)& =&  \intpt \left(\udp(\vx,t) \, a_\vp + \vdp^*(\vx,t) \, b^\dagger_\vp\right)\label{field}\, ,\\
\udp(\vx,t) & = & \sqrt\frac{\ena}{\eb\left(\ena+\eb\right)} e^{-i(\ena t- \vp \cdot \vx)}\label{possol}\, ,\\
\vdp(\vx,t) & = & \sqrt\frac{\eb}{\ena\left(\ena+\eb\right)} e^{-i(\eb t- \vp \cdot \vx )}\label{negsol} \, .
\end{eqnarray}

Equations (\ref{ham1}) and (\ref{com1}) describe a theory of particles with energy-momentum relation
\be
\frac{\ena(\vp)}{\omp}=\sqrt{1+\left(\frac{\theta \omp}2\right)^2}+\frac{\theta \omp}2 \label{disp_a}\, ,
\ee
and their antiparticles, with energy-momentum relation
\be
\frac{\eb(\vp)}{\omp}=\sqrt{1+\left(\frac{\theta \omp}2\right)^2}-\frac{\theta \omp}2 \label{disp_b}\, ,
\ee
where $\omp = \sqrt{\vp^2 + m^2}$. The difference of energies between particles and antiparticles with the same momentum is a consequence of the CPT violation of the theory.
Notice also that  $\ena(-\theta) =\eb (\theta)$, as it should be, due to the symmetry $a\leftrightarrow b$, $\Phi\leftrightarrow \Phimas$, $\theta \rightarrow -\theta$. Therefore $\vdp(\vx,t;\theta)=\udp(\vx,t;-\theta)$.

The equation of motion of the field can be obtained by using the Heisenberg equation for the evolution of operators in the Heisenberg picture. The result is
\be
L(x,\partial) \Phi(x)=\left[\partial_0^2+(m^2-\Delta)(1+ i \theta \partial_0)\right] \Phi(x) = 0 \label{motion1}\, .
\ee
If we define the following internal product in the space of solutions of (\ref{motion1}),
\begin{eqnarray}
\left(\varphi_1(\theta),\varphi_2(\theta)\right) & = &  -i\intxt
 \left(\varphi_1(\theta)\frac {\textstyle 1}{\textstyle 1-i\theta\partial_0}\partial_0 \varphi_2^*(\theta) \right.\nonumber \\
& & \left. -  \varphi_2^*(-\theta)\frac {\textstyle 1}{\textstyle 1-i\theta\partial_0}\partial_0 \varphi_1(-\theta)
 \right)\, ,
\label{prod}
\end{eqnarray}
then $\udp(\vx,t)$ and $\vdp^*(\vx,t)$ form an orthonormal (in the sense of Ref.~\cite{BD}) basis of solutions of the equation of motion.

As a side remark let us notice that equations (\ref{disp_a}), (\ref{disp_b}), (\ref{motion1}), show for $m=0$ a striking similarity with the dispersion relation
\be
\omega = \pm \sqrt{c^2 k^2 -\left(\frac{2 \nu k^2}{3} \right)^2} -i \frac{2\nu k^2}3\, ,
\ee
 and equation of motion
\be
\partial_t^2 \psi_1 \, = \, c^2\bm{\nabla}^2 \psi_1 + \frac 43 \nu \partial_t \bm{\nabla}^2 \psi_1
\ee
of sound waves propagating through a viscous fluid of viscosity $\nu$ at rest ($\vec{v}_0 = 0$) \cite{Visser:1997ux}. They turn out to be the same, taking into account that the frequency associated to a b-mode of energy $\eb$ is $-\eb$, and redefining $i\theta = \frac {4}{3} \nu$.
The meaning of a purely imaginary viscosity is far from clear, although the main consequence is obviously that the dissipative effects associated to viscosity are turned into dispersive ones, the dispersion relation becoming real.

\section{Symmetries of QNCFT in Flat Spacetime}

A symmetry of a theory is defined as a transformation which leaves the action of the theory invariant. A symmetry also has the property of transforming solutions of the equation of motion into solutions of the equations of motion. Thus, if we want to search for the symmetries of the theory, a good starting point would be to find the operators $\mathcal{O}(x,\partial,...)$ such that if $\Phi(x)$ is a solution of \eqref{motion1}, then also $\mathcal{O}(x,\partial,...)\Phi(x)$ is. {This implies that the action of the commutator of $L$ and $\mathcal{O}$ on solutions of the field equations has to vanish}
\be
\left[L(x,\partial,...),\mathcal{O}(x,\partial,...)\right]\Phi(x) = 0 \label{prop}\, .
\ee
The above program is carried out more easily if one Fourier transforms the field $\Phi(x)=\intpc e^{-i p\cdot x}\Phi(p)$.
Then, the equation of motion turns out to be \cite{MasterDiego},
\be
L \Phi(p) = \left\{ -p_0^2+(m^2+\vp^2)(1+\theta p_0) \right\}\Phi(p) = 0\, ,
\label{eq:disp}
\ee
and the operators  verifying the property (\ref{prop}) are only
\begin{eqnarray}
\mathcal{P}_i & = & p_i \, , \label{eq:sy1}\\
\mathcal{P}_0 & = & p_0\, , \label{eq:sy2}\\
\mathcal{M}_{ij} & = & p_i \frac \partial {\partial p^j}-p_j \frac \partial {\partial p^i}\, , \label{eq:sy3}\\
\mathcal{M}_{0i} & = & \frac {p_0}{\sqrt{1+\theta p_0}} \frac \partial {\partial p^i}-p_i \frac{(1+\theta p_0)^{3/2}}{1+\theta p_0/2} \frac \partial {\partial p^0}\, , \label{eq:sy4}
\end{eqnarray}
or arbitrary functions of them. Therefore the previous operators can be considered the generators of the symmetry group.

It is now evident that while the generators of translations $\mathcal{P}_i ,\mathcal{P}_0$ and rotations $\mathcal{M}_{ij}$ are left unchanged, a new set of generators for the boosts $\mathcal{M}_{0i}$ is induced by the noncanonical commutation of the field.
Notice however, that if one  restricts $p_0$ so that $p_0>-1/\theta$, then under the change of variables
\be
\tilde{p}_0 = \frac {p_0}{\sqrt{1+\theta p_0}}\, , \qquad \tilde{p}_i = p_i \, ,\label {aux}
\ee
the generators of the symmetry group can be rewritten as:
\begin{eqnarray}
\mathcal{P}_i & = & \tilde{p}_i \, , \\
\mathcal{P}_0 & = & \tilde{p}_0\left(\sqrt{1+\theta^2 \tilde{p}_0^2/4}+ \theta \tilde{p}_0/2 \right)\, ,\\
\mathcal{M}_{ij} & = & \tilde{p}_i \frac \partial {\partial \tilde{p}^j}-\tilde{p}_j \frac \partial {\partial \tilde{p}^i}\, ,\\
\mathcal{M}_{0i} & = & \tilde {p}_0 \frac \partial {\partial \tilde{p}^i}-\tilde{p}_i \frac \partial {\partial \tilde{p}^0}\, \label{gens}.
\end{eqnarray}

This shows that the generators of rotations and deformed boosts are isomorphic to the generators of the Lorentz group and it is only the action of the boosts on the four-momentum that it is deformed due to the presence of $\theta$ (reducing to its standard form in the limit of $\theta\to 0$). This is in striking similarity with what is normally conjectured in DSR scenarios \cite{AmelinoCamelia:2000mn, AmelinoCamelia:2002wr, Judes:2002bw} \footnote{Note that in the DSR literature as well as in Ref.~\cite{NCFisDSR} greek letters $(\pi_0,\bm{\pi})$ are used to refer to the auxiliary variables $(\tilde{p}_0,\tilde{p}_i)$. We chose here a different notation in order to render the expressions in the rest of the paper more understandable. } and hence it is no surprise that also in this case the whole group of symmetries can be made isomorphic to the Poincar\'e group when $\mathcal{P}_0$ is substituted by $\tilde{\mathcal{P}}_0 = \tilde{p}_0$, which is also the generator of a symmetry. This clearly hints at the possibility to recast the theory in an auxiliary spacetime associated with momenta $(\tilde{p}_i ,\tilde{p}_0)$ in which the action of the Poincar\'e group will be standard. We shall explore this possibility in what follows but before a few remarks about the nature of the just found symmetry are in order.

 Symmetries in field theories can be of two types. Internal symmetries are transformations in the field space which do not mix fields at different spacetime points. These transformations will have the form $\Phi(x)\rightarrow\Phi'(x)$ with $\Phi'(x)=\mathcal{O}(x)\Phi(x)$. We will call instead external symmetries those symmetries which do mix fields at different spacetime points. For example this can happen if the transformation affects also spacetime itself. Indeed, spacetime symmetries are transformations in the spacetime background which affect the field. These transformations will have the form $x\rightarrow x'(x)$ and $\Phi(x)\rightarrow \Phi'(x')$ where, in the case of the scalar field, $\Phi'(x')=\Phi(x)$. Therefore spacetime symmetries are external symmetries. However, the opposite is not always true. In particular the symmetry associated to (\ref{eq:sy1}), (\ref{eq:sy2}), (\ref{eq:sy3}), (\ref{eq:sy4}),  is not a proper spacetime symmetry because of its non-local nature.

 Coming back to our study of the symmetries of the theory in flat spacetime we now want to explore the possibility to recast the whole field theory in a auxiliary spacetime in which the Poincar\'e group acts in the usual way. We shall do so by looking for transformations which leave the action invariant. While a rigorous treatment can be found in Ref.~\cite{NCFisDSR} we shall provide here a more concise derivation for the sake of simplicity.

The first step would be to find the action $S$, which will be the spacetime integral of some Lagrangian density $\Lag$. In order to achieve this, we will write (\ref{ham1}) in terms of field variables whose commutation relations are canonical,
\be
\Phi_c = \Phi-\frac{i\theta}2 \Pi \label{phic}\, .
\ee
Then, writing the Hamiltonian in these variables and performing the canonical transformation
\begin{eqnarray}
\bar{\Phi}  & = & \left(1-\frac{i\theta}2 \sqrt{m^2-\Delta}\right)^{-1}\Phi_c \, ,\nonumber \\
\bar{\Pi} & = & \left(1-\frac{i\theta}2 \sqrt{m^2-\Delta}\right) \Pi \nonumber \, ,
\end{eqnarray}
the action can be written, up to boundary terms, as
\begin{eqnarray}
S & = & \intxc \Lag (x) \nonumber \\
& = & \intxc \left(\Pimas \dot{\Phi}_c + \dot{\Phimas}_c \Pi - \Ham (\Phi(\Phi_c),\Pi)\right) \nonumber \\
& = & \intxc \left(\bar{\Pi}^\dagger \dot{\bar{\Phi}} + \dot{\bar{\Phi}}^\dagger \bar{\Pi} - \Ham (\Phi(\bar{\Phi}),\Pi(\bar{\Pi}))\right) \nonumber \\
& = & \intxc \, \bar\Phi^\dagger \left( -\partial_0^2-(1+i\theta\partial_0) (m^2-\Delta)\right)\bar{\Phi} \label{lag1} \, ,
\end{eqnarray}
where in the last line we have substituted the momentum as a function of the field and its derivatives via the Hamilton-Jacobi equations.

 We want now to show that the action \eqref{lag1} can be expressed as a standard scalar field action in flat spacetime. We can start by introducing the field redefinition
\be
\phi(x) = \sqrt{1+i \theta \partial_0}\bar{\Phi}(x)\, ,
\ee
 so that the action takes the simpler form,
\be
S = \intxc \, \phi^\dagger(x) \left( \frac{-\partial_0^2}{1+i\theta\partial_0}-(m^2-\Delta)\right)\phi (x)\, .
\ee
Let us look at this action in momentum space by using the Fourier transform of $\phi(x)$, $\phi(p)$
\be
S = \intpc \phi^\dagger(p) \left( \frac{p_0^2}{1+\theta p_0}-(m^2+\vp^2)\right)\phi (p)\, \label{lagfrac},
\ee
where the integral over $p_0$ goes from $-\infty$ to $+\infty$. Now we can split the action into two terms $S^\theta$ and $\bar{S}^\theta$, with the integral over $p_0$ ranging respectively from $-1/\theta$ to $+\infty$ and from $-\infty$ to $-1/\theta$. All the solutions of the classical equations of motion lay in $S^\theta$, but the $\bar{S}^\theta$ term can affect the quantum behavior of the theory~\cite{NCFisDSR}. Changing in $S^\theta$ the integration variable $p_0$ for $\tilde{p}_0$, defined in (\ref{aux}),  this can be rewritten as
\be
S^\theta = \intpt \frac{d\tilde{p}_0}{2 \pi}  \frac{\partial p_0}{\partial \tilde{p}_0}\phi^\dagger(p(\tilde{p})) \left( \tilde{p}_0^2-(m^2+\vp^2)\right)\phi (p(\tilde{p}))\, .
\ee
 Finally we can define the auxiliary field
 \be
 \tilde{\phi}(\tilde{p})=\sqrt{\frac{\partial p_0}{\partial \tilde{p}_0}}\,\phi (p(\tilde{p})) \, ,
\label{auxfield}
 \ee
 and inverse Fourier transform, so to arrive to the standard Klein-Gordon-action in some ``auxiliary spacetime".
\be
S^\theta =\intxt \, d\tilde{x}_0\, \tilde{\phi}^\dagger (\tilde{x})\left(-\eta^{\mu\nu}\partial_\mu\partial_\nu - m^2 \right) \tilde\phi(\tilde{x})\, . \label{actaux}
\ee

Now the properties of the symmetry group of the theory can be understood. When written in the auxiliary variables $(\tilde{x}_0,\vx)$ (auxiliary frame), the symmetry group is the Poincar\'e one associated to the flat auxiliary spacetime symmetries of $S^\theta$.
Reversely, if we assume that the Poincar\'e group acts trivially on $\bar{S}^\theta$, its action on the real spacetime and the original field variables (original frame) can be derived by pulling back the operations that we have applied above. In this case the action of the group is found to be realized in some non-local way. Indeed, this setting is similar to the one proposed in the DSR framework~\cite{AmelinoCamelia:2000mn}.\footnote{Note that the above treatment does not depend on the detailed form of the field dispersion relation nor on its origin. In principle, the discussion may be applied to more general dispersion relations.}

However, it has to be noticed that the necessary splitting of the action into two terms implies that our initial theory is not exactly equivalent to a relativistic field theory in flat space, at the quantum level \cite{NCFisDSR}. As a matter of fact, one might argue that the symmetry we have found is a fake one, because it invokes non-locality (see e.g.\/ the discussion about symmetries of Maxwell equations in \cite{Giulini:2006uy}).

Indeed spacetime symmetries are local because the transformed field in a spacetime point is a function of the value of the field in just one spacetime point. In the case we are considering, the $\mathcal{M}_{0i}$ are generating a family of external transformations which cannot be interpreted as spacetime symmetries, at least in the usual way, due to their intrinsic non-locality. Indeed such a symmetry transforms a whole field configuration, which is a solution of the equation of motion, into a different field configuration which is also a solution.

Finally, one must also take into account that the Hamiltonian of the theory, the space integral of Eq.~(\ref{ham1}),  does not belong to the set of generators of this Poincar\'e-like symmetry group, although time translations are a symmetry of the free theory. The problem is that the commutator of the Hamiltonian with the generator of boosts would end up defining a new generator and hence the algebra would not be not closed.

 An infinite set of generators of the symmetry group of the free theory in flat spacetime
may be obtained by iteration of this procedure. However, the whole symmetry group forbids
the addition of new terms in the action or the appearance of a spacetime dependence of
the coefficients of the existing terms. Therefore the symmetry group must be reduced
(broken into subgroups) in order to extend the theory.

These considerations force us to take a decision. We either try to generalize the theory to curved spacetime by promoting the non-local implementation of the Poincar\'e group to a gauge symmetry, and therefore time translations are taken as an accidental symmetry of the theory in flat spacetime  (or at low energies where $\theta$ corrections are supposed to be negligible), or we can take an alternative point of view and consider as the physical symmetries only those realized in a local way while boost invariance is broken.

The first approach has been shown to suffice in the construction of a renormalizable interacting theory in flat spacetime~\cite{NCFisDSR}, but, as previously discussed, some of the intuitive notions of what a symmetry is can be spoiled by the non-locality of the implementation. While more intuitive, the second approach will imply a preferred time direction and preferred slices of simultaneity in spacetime.

\section{QNCFT in Curved Spacetimes}

The aforementioned alternative points of view regarding the actual symmetries  imply that  two different approaches in constructing a QNCFT in Curved spacetimes can be taken depending on which is the symmetry of the free field theory in flat spacetime we would like to promote to a gauge symmetry of the spacetime. Hence, we will perform them separately in the following subsections.

\subsection{A spacetime with a gauge DSR-like Invariance?}

One of the ways of formulating General Relativity (GR) consists in turning the global Poincar\'e symmetry of Minkowski spacetime into a gauge symmetry of any curved geometry.\footnote{ Indeed the Strong Equivalence Principle (which selects GR among relativistic theories of gravitation with spin-2 gravitons) entails Local Lorentz invariance as a fundamental postulate (together with universality of free fall and local position invariance of experiments, including gravitational ones)\cite{Will}.} This procedure provides also a link between Quantum Field Theory (QFT) in flat spacetimes and QFT in curved backgrounds. We will try to follow the same steps by gauging the symmetry we have found for the noncanonical field.

A global symmetry in the noncanonical field has been found, whose generators are:
\begin{eqnarray}
\mathcal{P}_i & = & -i\partial_i \, ,\nonumber\\
\mathcal{\tilde{P}}_0 & = & i\frac{\partial_0}{\sqrt{1+i\theta\partial_0}} \, ,\nonumber\\
\mathcal{M}_{ij} & = & x_i\partial_j - x_j\partial_i \, ,\nonumber\\
\mathcal{M}_{0i} & = & x_i\frac{\partial_0}{\sqrt{1+i\theta\partial_0}}-x_0\partial_i \frac{(1+i\theta\partial_0)^{3/2}}{1+i\theta\partial_0/2} \label{gen}\, .
\end{eqnarray}
These operators satisfy the Poincar\'e Algebra. We want to generalize our considerations about the symmetry to curved spacetime without loosing this connection. We can do this by introducing the tetrad field formalism  i.e.\/ by constructing a set of normal coordinates $y^a_X$ at each point $X$ of the spacetime manifold (for example for time-like observers one can pick up as a preferred axis the one associated to the tangent vector to the observer worldline and the other three axis as an orthogonal basis in the spacelike hypersurface orthogonal to such vector).

The relationship between these coordinates and the general coordinates in the manifold defines the tetrad field $e^a_\mu (x) = \frac{\partial y^a_X}{\partial x^\mu}\vert_{x=X}$.  In terms of the $y^a_X$ the metric at $X$ is simply $\eta_{ab}$ (which is used to lower latin indices), and can be transformed to the metric expressed in terms of the general coordinates $g_{\mu\nu}(x)$ (which is used to lower greek indices) with the use of the tetrad.

In the standard case, we must introduce the covariant derivative, an operator which turns global Poincar\'e and Lorentz invariant tensors into local Poincar\'e and general coordinate transformations  invariant tensor.
\be
D_\mu = \partial_\mu - e^a_\mu (x) \mathcal{P}_a -\frac 12 \omega^{ab}_\mu (x) \mathcal{M}_{ab} \label{covar}\, ,
\ee
where $e^a_\mu (x)$ and $\omega^{ab}_\mu (x)$ (the spin connection) play the role of gauge fields.  This leads from QFT in Flat Spacetime to QFT in Curved Spacetimes when partial derivatives $\partial_a$ are replaced by covariant derivatives $D_a = e_a^\mu D_\mu$ and Lorentz tensors are replaced by generalized tensors built up with the tetrad.

However, if this procedure is applied to the case of the noncanonical field, changing the partial derivatives by covariant derivatives in the action (\ref{lag1}) and inserting the tetrad field does not make the gauge transformation (\ref{gen}) a gauge symmetry. The connection between the QNCFT and the relativistic QFT (\ref{actaux}) is broken.

 In order to see this, let us try to follow the same procedure as in the standard case. The action will have the form
\begin{eqnarray}
S & = & \intxc \sqrt{-g(x)} \left[\Pimas e_0^\mu(x) (D_\mu \Phi) + e_0^\mu(x) (D_\mu \Phimas) \Pi \right.\nonumber\\
&&- \Pimas\Pi - \delta^{ij}e_i^\mu(x) (D_\mu \Phimas) e_j^\nu(x) (D_\nu \Phi) - m^2\Phimas\Phi\nonumber\\
&&\left.  -\frac{i\theta}{2} \left(\Pimas e_0^\mu(x) (D_\mu \Pi) - e_0^\mu(x) (D_\mu \Pimas) \Pi \right) \right].\label{eq:break}
\end{eqnarray}
This action is explicitly general covariant and in the flat spacetime limit does reproduce the second line of Eq.~(\ref{lag1}). However, it is easy to see that there is no transformation enabling one to write \eqref{eq:break} as the curved spacetime generalization of the last line of Eq.~(\ref{lag1}) due to the explicit dependence of the tetrad field on the spacetime coordinates. Indeed, the derivation of Eq.~(\ref{lag1}), required infinite integrations by parts. If these were done in the curved spacetime case, an infinite series of new terms would appear in the action, involving covariant derivatives of the tetrad fields. Furthermore, it is also possible to check that the infinitesimal gauge transformations generated either by $\mathcal{\tilde{P}}_0$ or by $\mathcal{M}_{0i}$ do not leave the Lagrangian associated with \eqref{eq:break} invariant even at linear order in $\theta$.

These evident problems lead us to the conclusion that there seems to be a fundamental obstruction in gauging external symmetries generated by non-local operators and in particular in making the symmetry generated by (\ref{gen}) a gauge symmetry of the field action.

However, in the previous section we have shown that the non-locally implemented symmetry can be seen as a proper (i.e.~local) spacetime symmetry of a new auxiliary $\tilde{x}$-spacetime. Though non-locality prevents the gauging of symmetries in $x$-spacetime, the procedure of gauging spacetime symmetries and its results are well known. Therefore we can argue that the proper way to proceed is to gauge global DSR-like symmetries in $\tilde{x}$-spacetime. Of course, the procedure of $\tilde{x}$-gauging the global Poincar\'e symmetry leads to the standard QFT in curved spacetime, but now the curved spacetime is the auxiliary one.

In those regions in which the auxiliary spacetime is asymptotically flat, the mapping between auxiliary and noncanonical fields (\ref{auxfield}) can be used and the propagator in the  noncanonical field, in physical spacetime, can be found. However, a connection between the auxiliary field and the noncanonical field in a point of auxiliary spacetime in which the metric is not asymptotically flat seems precluded, at least in the present treatment.

Let us then reformulate all the theory in the auxiliary frame, in which everything is standard. The {part of the action which contains the solutions of the equations of motion} in this DSR-like approach would be
\be
S^\theta=\intauxx \gaux \tilde{\phi}(\tilde{x})\left(-\tilde{g}^{\mu\nu}\tilde{\nabla}_\mu\tilde{\nabla}_\nu - m^2 +\xi \tilde{R}(\tilde{x})  \right) \tilde\phi(\tilde{x})\, ,
\ee
where $\tilde{\nabla}$ means covariant derivative, $\tilde{R}(\tilde{x})$ is the Ricci scalar, and the line element in the $\tilde{x}$-spacetime is given by
\be
d\tilde{s}^2 = \tilde{g}_{\mu\nu} d\tilde{x}_\mu d\tilde{x}_\nu \nonumber\, .
\ee

The theory in the auxiliary variables is ordinary QFT in Curved Spacetime \cite{BD}. The equation of motion of the field in auxiliary spacetime is,
\be
\left(\tilde{g}^{\mu\nu}\tilde{\nabla}_\mu\tilde{\nabla}_\nu + m^2 +\xi \tilde{R}(\tilde{x}) \right) \tilde\phi(\tilde{x})=0 \label{mov2}\, .
\ee
Its solutions are spanned by a basis $\left\{\tilde{u}_\vp(\tilde{x}),\tilde{u}^*_\vp(\tilde{x})\right\}$ which is orthonormal under the internal product
\be
\left(\varphi_1,\varphi_2\right)=  -i\int_{\tilde{\Sigma}} d\tilde{\Sigma}\sqrt{-\tilde{g}_{\tilde{\Sigma}}(\tilde{x})}\tilde{n}_\mu
\left(\varphi_1 \tilde{\partial}_\mu \varphi_2^*-\varphi_2^*\tilde{\partial}_\mu \varphi_1 \right)\, ,
\ee
where $\tilde{\Sigma}$ is some spacelike hypersurface, $\tilde{n}$ the future-oriented timelike vector orthonormal to it. The value of the internal product is independent of the choice of $\tilde{\Sigma}$.

From the above discussion the standard approach to QFT in curved spacetime straightforwardly follows. In particular standard problems like particle creation from the vacuum can be approached with the usual Bogoliubov techniques. Of course, one may wonder how far one could trust such calculations given that the correspondence between the original, physical, spacetime and the auxiliary one (in which one ends up working) is only via the $S^\theta$ part of the real action. In this sense it is important to notice that the virtual modes belonging to the $\bar{S}^\theta$ term of the action do not affect the Bogoliuvov transformation between modes in the $S^\theta$ term, as these virtual modes ``do not see" the auxiliary spacetime.

The conclusion is that working in this framework the resulting theory will give then the same results as the standard QFT in curved spacetime, at least insofar one is working in the auxiliary spacetime. It still remains open the issue of  mapping back the results to the original spacetime which, as we said, seems doable only in asymptotically flat regions and hence not generically.

\subsection{Foliating spacetime with noncanonical commutation relations}

Let us adopt the alternative point of view that meaningful symmetries are local implementations of a certain symmetry group, in this case rotations in space and translations in space and time, {and that the DSR-like Poincar\'e group is just an accidental symmetry of the free action in flat spacetime}. Therefore, we would like to curve physical spacetime in a way such that just spatial rotations and spacetime translations are preserved. It should be clear that in this case, a preferred arrow of time emerges in which the fields appearing in the commutation relations are simultaneous. Each set of simultaneous events defines a spacelike hypersurface in the manifold of spacetime, foliating it, while the orthogonal, timelike direction defines the evolution in time. In those hypersurfaces no spacelike coordinates are preferred, and thus the commutation relations of the field in flat spacetime (\ref{com1}) have to be  suitably generalized in order to make them covariant under general coordinate transformations on the spacelike hypersurfaces,
\begin{eqnarray}
\left[\Phi(\vx,t),\Phimas(\vx',t)\right]& =&\frac \theta{\sqrt{h(t)}} \deltat(\vx-\vx') \nonumber\\
\left[\Phi(\vx,t),\Pimas(\vx',t)\right]&=&\frac {i\hbar}{\sqrt{h(t)}} \deltat(\vx-\vx') \nonumber\\
\left[\Pi(\vx,t),\Pimas(\vx',t)\right]&=&0\,
\end{eqnarray}
where $h(t)$ is the determinant of the three-metric $h_{ij}(t)$ associated to each spacelike hypersurface of simultaneity. This foliation of spacetime into slices of simultaneity naturally leads to the Arnowitt Deser Misner description of GR (ADM) \cite{Arnowitt:1960es, Arnowitt:1962hi}.
 We briefly review it here for completeness and for fixing the notation.

In the ADM formalism one takes a foliation of spacetime in spacelike hypersurfaces, which implies a split of the whole spacetime metric, the four-metric, into its spacelike-spacelike components $g_{ij}(\vx,t)$ --- which coincide in this frame of reference with the ones of the induced metric on the hypersurfaces (the three-metric $h_{ij}$) --- a three-vector shift function $N^i(\vx,t)$ and a three-scalar lapse function $N(\vx,t)$. The components of the four-metric in this frame are then
\be
\begin{array}{c}
g_{\mu\nu}
\end{array}
=
\left(
\begin{array}{cc}
g_{00} & g_{0j}\\
g_{i0} & g_{ij}
\end{array}
\right)
=
\left(
\begin{array}{cc}
N^2-N^i N^k h_{ik} & -h_{kj} N^k \\
-h_{ik} N^k & -h_{ij}
\end{array}
\right)\, .\label{ADM}
\ee
It has to be noticed that the formalism is still covariant under general spacelike coordinate transformations. Spacelike indices $i$,$j$,$k$,... are lowered with the three-metric $h_{ij}$ and raised with the inverse of the three-metric, $h^{ij}$. The inverse of the four-metric turns out to be
\be
\begin{array}{c}
g^{\mu\nu}
\end{array}
=
\left(
\begin{array}{cc}
1/N^2 & -N^j/N^2 \\
-N^i/N^2 & -h^{ij}+N^i N^j/N^2
\end{array}
\right)\, ,
\ee
and the determinant of the four-metric is $g = -N^2 h$. The {timelike} vector which is orthonormal to the spacelike hypersurfaces is
\be
n^\mu = (1/N, -N^i/N) \, . \label{time}
\ee

The covariant four derivative is constructed in the usual way with the affine connections which are furthermore assumed to be the Christoffel symbols $\Gamma^{(4)\lambda}_{\mu\nu}$ associated with the four-metric $g_{\mu\nu}$.
The extrinsic curvature $K_{ij}$ is defined with the covariant four-derivative of the normal to the hypersurfaces ($K_{ij}= N \Gamma^{(4)0}_{ij})$. The covariant three-derivative is defined as the projection of the covariant four-derivative on the spacelike hypersurface, $\nabla^{(3)}_i V^j \equiv V^j_{\vert i} = \partial_i V^j + \Gamma^{(3)j}_{ik}V^k$, where $\Gamma^{(3)j}_{ik} = \Gamma^{(4)j}_{ik} + K_{ik}N^j/N$ coincides with the Christoffel symbols built up with the three-metric. (From now on the superscript $(3)$ can be dropped in order to simplify the notation.)  The intrinsic curvature $R^i_{jkl}$ is built in the usual way with the three-metric $h_{ij}$ and its derivatives, and can be related to the four-curvature and the extrinsic curvature.

Finally, this notation has also a link to the tetrad notation introduced in the previous subsection. The tetrad $e_a^\mu$ associated with the metric (\ref{ADM}) is
\be
\begin{array}{ccc}
e_0^0 \, = \, 1/N & & e_0^i \, = \,N^i/N \\
e_i^0 \, = \, 0 & & e_j^i \, = \,e^{(3)j}_i \\
\end{array}\, ,\label{tetradADM}
\ee
where $e^{(3)i}_a$ is the dreibein associated to the $h_{ij}$ three-metric.

We have now to consider how can we couple the noncanonical field to the physical spacetime metric (\ref{ADM}). The result was already given in the previous subsection in Eq.~(\ref{eq:break}). Written in the notation of the ADM prescription with the use of (\ref{tetradADM}), the expression of the action, Lagrangian density and Hamiltonian density of the field are
\begin{eqnarray}
S_\Phi & = & {\textstyle \intxc }\sqrt{h} \Lag_\Phi \, ,\label{actionADM}\\
\Lag_\Phi & = & \Pimas\dot{\Phi}_c + \dot{\Phi}^\dagger_c \Pi-N \Ham_\Phi \, , \label{lagADM} \\
\Ham_\Phi & = & \Pimas \Pi+h^{ij}\partial_i\Phimas\partial_j\Phi+m^2\Phimas\Phi \nonumber\\
&& +\left(\frac{N^i}N \Pimas \partial_i\Phi_c \,+\, h.c.\,\right)\, . \label{hamADM}
\end{eqnarray}
where $\Phi_c$ is given by (\ref{phic}). The stress energy tensor is defined as
\be
T_{\mu\nu}  =  \frac 2{N\sqrt{h}}\frac{\delta S_\Phi}{\delta g^{\mu\nu}}\, ,
\ee
and turns out to be
\begin{eqnarray}
T_{00} & = & N\Pimas\dot{\Phi}_c +N N^i\Pimas \partial_i\Phi_c+N^i N^j \partial_i \Phimas \partial_j \Phi\, +\,h.c.\nonumber\\
& & -(N^2-N_i N^i)\Lag_\Phi \nonumber\\
T_{0i} & = & N^j\partial_i\Phimas \partial_j\Phi+N\Pimas\partial_i\Phi_c \,+\,h.c.\,+N_i \Lag_\Phi \nonumber\\
T_{ij} & = & \partial_i\Phimas \partial_j \Phi \, +\, h.c.\, +h_{ij} \Lag_\phi\, .
\end{eqnarray}

Now we have all the information required about the coupling of the noncanonical field to the gravitational potential in curved physical spacetime. In particular, we can derive from the action (\ref{actionADM}) the equation of motion of the field. These equations can be rewritten in a more compact way if we write them in terms of the Lie derivative of the fields along the timelike direction (\ref{time}),
\be
 \Lie\Phi = \frac 1 N \dot{\Phi} - \frac{N^i} N \partial_i\Phi\, .
\ee
Varying the action with respect to $\Pimas$ we get
\be
N\Lie\Phi-i\theta N\Lie\Pi-N\Pi-\frac{i\theta}4 \frac{\dot h}h\Pi + \frac{i\theta}2 N^i_{\vert i}\Pi\,=\, 0 \label{PiADM}
\ee
and varying the action with respect to $\Phimas$,
\be
-N\Lie\Pi-\frac{\dot{h}}{2 h}\Pi-N m^2\Phi+(N \partial_i \Phi)^{\vert i}+N^i_{\vert i}\Pi\,=\, 0\, .\label{PhiADM}
\ee
These equations define a system of coupled differential equations which should be solved provided the metric of spacetime is known and we neglect the back-reaction of the field on the metric. We should remark that the general covariance of the whole spacetime is broken by the preferred choice of a time variable, but the general covariance on the spacelike hypersurfaces is guaranteed by construction. This setting reminds the context of analogue gravity models \cite{Barcelo:2005fc}.

{The formulation of the quantum theory of a field in curved spacetime requires the definition of a scalar product in the space of solutions of the field equations. We are now able to define an inner product in the space of solutions. The inner product essentially carries the information of the commutation relations. Given two solutions, $\Phi = \varphi_{1,\theta}$ and  $\Phi = \varphi_{2,\theta}$, of the  equations of the field (\ref{PiADM}), (\ref{PhiADM}) we} can define the internal product $\left(\, , \,\right)$ as
\be
\left(\varphi_1, \varphi_2\right)  =  -i\intxt \sqrt{h} \left[\varphi_{1,\theta}\varphi^{\Pi\,*}_{2,\theta}-\varphi^*_{2,-\theta}\varphi^\Pi_{1,-\theta}\right]\, ,
\label{intprod}
\ee
where the integral is evaluated at any of the spacelike hypersurfaces in which spacetime becomes foliated and
\be
\varphi_{r,\theta}^\Pi  =  \left[ 1+\frac{i\theta}N\left(N\Lie+\frac{\dot{h}}{4h}+\frac{N^i_{\vert i}}{2}\right) \right]^{-1} \Lie\varphi_r (\theta)\, ,
\ee
is the conjugate momentum solution $\Pi = \varphi_{r,\theta}^\Pi$ associated to the field solution $\Phi = \varphi_{r,\theta}$ with $r=1,2$. This internal product preserves the symplectic structure of the theory and is independent of the choice of the hypersurface by construction \cite{Wald}.

Let $\{\Phi \, = \, \udp(\vx,t;\theta)\}$ be an orthonormal basis of solutions of the system of differential  equations (\ref{PiADM}), (\ref{PhiADM}) with positive frequency with respect to the preferred time direction (\ref{time}), labeled by the index $\vp$. Let $\{\Phi \, = \, \vdp^*\}$ be an orthonormal basis of solutions of negative frequency of the same system of differential equations. Then the field can be expanded in terms of annihilation operators of the $\udp(\vx,t;\theta)$ and creation operators of the $\vdp^*(\vx,t;\theta)$
\be
\Phi(\vx,t)=\intpt\left[\udp(\vx,t;\theta) \,a_\vp\,+\,\vdp(\vx,t;\theta)^* \,b^\dagger_\vp\right]\, .
\ee
 If $\vdp^*$ is a solution of negative frequency of the system of field equations (\ref{PiADM}), (\ref{PhiADM}), then $\vdp$ is a solution of positive frequency of the complex conjugate of the system, which coincides with the result of changing $\theta\,\rightarrow\, -\theta$ in the system. Therefore $\vdp(\vx,t;\theta) \propto \udp (\vx,t;-\theta)$. A more careful analysis of the commutation relations shows that $\vdp(\vx,t;\theta) = \udp (\vx,t;-\theta)$. This is  due to the symmetry $\theta\,\rightarrow \,-\theta, \, a\,\leftrightarrow \, b, \,\Phi \,\leftrightarrow \, \Phimas, \, \Pi\,\leftrightarrow \,\Pimas $.

As usual, there is non-uniqueness in the choice of an orthonormal basis $\{\udp,\vdp^*\}$ of solutions of the system of field equations if the metric induces a loss of the time translation symmetry. We must then resort to Bogoliubov techniques in order to relate the associated non equivalent vacua. Let us expand the field in terms of two different basis of solutions of the equation of motion
\begin{eqnarray}
\Phi(x) & = & \intpt \left[ \udp(x)\,  a_\vp +\vdp^*(x) \, b^\dagger_\vp \right]\nonumber\\
& = & \int \frac{d^3 \vp'}{(2\pi)^3} \left[ \bar{U}_{\vp'}(x) \, \bar{a}_{\vp'} +\bar{V}_{\vp'}^*(x) \, \bar{b}_{\vp'}^\dagger \right]
\label{expansion}\, .
\end{eqnarray}
As both sets of solutions of the equation of motion are basis, we can define the Bogoliuvov transformation as a change of basis in the space of solutions of the equation of motion,
\be
\udp =  \int \frac{d^3 \vp'}{(2\pi)^3} \left[ \bar{U}_{\vp'}(x) \alpha_{\vp' \vp} +\bar{V}_{\vp'}^*(x) \beta_{\vp' \vp} \right] \, ,\label{Bogo}
\ee
where
\begin{eqnarray}
\alpha_{\vp' \vp} =  \left( \udp , \bar{U}_{\vp'} \right)\, , \qquad \beta_{\vp' \vp} & = & -\left( \udp , \bar{V}^*_{\vp'} \right)
\, .
\end{eqnarray}
We define a state $\vert 0 \rangle$ as the vacuum in the $\{\udp,\vdp^*\}$ basis. We want to compute the expectation value in this state of the number operators $\bar{N}_{a,\vp'}$, $\bar{N}_{b,\vp'}$ defined in the $\{\bar{U}_{\vp'},\bar{V}_{\vp'}^*\}$ basis. Inserting (\ref{Bogo}) in (\ref{expansion}) in order to get the Bogoliuvov transform of the creation-annihilation operators and plugging them in the expression of the number operators, the result will be
\begin{eqnarray}
\langle 0 \vert \bar{N}_{a,\vp'} \vert 0 \rangle & = &  \int d^3 \vp \vert \beta_{\vp' \vp}(-\theta) \vert ^2\, ,\\
\langle 0 \vert \bar{N}_{b,\vp'} \vert 0 \rangle & = &  \int d^3 \vp \vert \beta_{\vp' \vp}(\theta) \vert ^2\, .
\end{eqnarray}
Noticeably, the internal product (\ref{intprod}) has two terms which differ not only in sign but also in magnitude, unlike in the standard QFT in curved spacetime or the QNCFT in Curved auxiliary Spacetime. This two terms will in general be complex and oscillating with different phases.  Hence, when computing the number of particles and antiparticles observed by a detector in the vacuum of some other detector (like for example in the case of Rindler observers in flat spacetime or static observers at infinity in a black hole spacetime), one then finds that the corresponding spectrum follows the pattern of interference of the terms in the Bogoliuvov coefficient $\beta$. This result is essential for example in the studying black hole evaporation by Hawking radiation, something which we plan to address elsewhere.

\section{conclusions}

Recent works \cite{NCFisDSR} have offered a mechanism to identify new symmetries in the theory of QNCFT \cite{Carmona:2003kh}. This mechanism goes beyond this particular example and can be applied in principle to more general theories with modified dispersion relations. The new symmetry transformations will be non-local, but the action can be rewritten in terms of an auxiliary spacetime in which the symmetry transformations are local and the action is Poincar\'e invariant. There is a link between the conjugate four-momenta in physical and auxiliary spacetimes, which is reminiscent of the DSR paradigm \cite{Judes:2002bw}. As a result a non-local implementation of the Poincar\'e group is found to be a symmetry group of the theory in flat spacetime. However, the generators of the symmetries of this group do not form a closed algebra with the  generator of the symmetry of translations in time. When the theory is tried to be extended, it is unclear which of the symmetries are to be kept and which will be treated as accidental. Some works \cite{Giulini:2006uy} do not give much credit to non-locally implemented symmetries. On the other hand, in the absence of the Poincar\'e symmetry, the possibility of constructing a perturbatively renormalizable interacting theory of the noncanonical field is also unclear \cite{NCFisDSR}.

Two different paths have been followed in order to derive a QNCFT in Curved spacetimes depending on the symmetries that are required to be gauged: either the DSR-like symmetries of the free theory, though some of its elements are implemented non-locally, or the group of symmetries of the free theory that are locally implemented, i.e.~translations and rotations.

We find that curving the physical spacetime breaks the DSR-like symmetry. This maybe due to an obstruction in gauging symmetries generated by non-local operators. Consequently, if one wants to preserve the DSR-like implementation of the Poincar\'e group, one must take the auxiliary spacetime as the one to be curved. This leads to a relativistic QFT in Curved Auxiliary Spacetime, which is very similar to the standard one. The correspondence between the fields in the auxiliary frame and fields in the physical spacetime can be trivially done in regions in which the auxiliary spacetime is asymptotically flat. Strictly speaking, the two theories are not really equivalent due to the presence of $\bar{S}^\theta$ part of the physical action. The role of the $\bar{S}^\theta$ modes, which cannot be mapped to auxiliary spacetime, is to fix time ordering in the original frame but they do not seem to have any observable influence on the auxiliary frame, for example they do not affect the Bogoliuvov transformations.

On the other hand, if only the symmetries that are implemented locally are required to characterize spacetime at short scales, the physical spacetime of the QNCFT can be made curve using the ADM prescription. The commutation relations must be adapted to fit in a general covariant frame. The action of the field coupled to the three-metric, lapse and shift functions can be written unambiguously. General coordinate invariance is broken by the commutation relations into invariance under general coordinate transformations on the slices of simultaneity, and reparameterizations of time. The resulting QNCFT in Curved Physical Spacetime seems to have the same structure as the ordinary one, but new phenomena related to the UV scale may appear. The main difference between this theory and the standard QFT in Curved spacetimes is that, due of the different energies of particles and antiparticles, the two terms appearing in the internal product (\ref{intprod}) become different not only in sign, but also in magnitude.

We conclude that the theories introduced offer clearly distinguishable outcomes for related phenomenology. This will be further studied in future work.

\acknowledgments

We would like to thank C.~Barcel\'o, J.~M.~Carmona, J.~L.~Cort\'es, D.~Maz\'on, J.~Rubio, L.~Sindoni and M.~Visser for fruitful discussions. J.I. would like to thank the Scuola Internazionale Superiore di Studi Avanzati (SISSA) for the hospitality during the development of this work. Financial support was provided by CICYT (project FPA2003-02948), DGIID-DGA (project 2008-E24/2). J.I. acknowledges an FPU grant from the Spanish Ministerio de Ciencia e Innovaci\'on (MICIIN).

\end{document}